\documentclass[aps,preprint]{revtex4}%
\usepackage{amsfonts}
\usepackage{amsmath}
\usepackage{amssymb}
\usepackage{graphicx}%
\providecommand{\U}[1]{\protect\rule{.1in}{.1in}}

\begin{document}
\preprint{\newpage}
\title[ ]{The Symmetry of Current of the Coherent State From the View of $O\left(
3\right)  $ $\sigma$-model}
\author{Xu-Guang Shi}
\affiliation{College of Science, Beijing Forestry University, Beijing 100083, P.R.China}
\author{Zhi-Jie Qin}
\affiliation{School of Physics and Engineering, Zhengzhou University, Zhengzhou 450001, P.R.China}
\author{}
\affiliation{}
\keywords{coherence; decoherence; topology, two-condensate superconductor}
\pacs{03.65.-w 02.40.-k}

\begin{abstract}
In this paper, we apply the reduced density trajectory, $\phi-$mapping
topological current theory and Ginzberg-Landau model to study the current of
the coherent state. We give the new expression of the current of the coherent
state. Based on this expression, the symmetry of the coherence is studied. We
find that the current of the coherent state corresponds to the supercurrent of
two-condensate system. The partial wave functions of the coherence carry new
charges and their interaction is mediated by new $U(1)$ gauge potential.\-

$\allowbreak$

\allowbreak

Keywords: coherence; decoherence; topology, two-condensate superconductor

PACS: 03.65.-w 02.40.-k

\-

\allowbreak The corresponding author's information

Email: shixg@bjfu.edu.cn

\end{abstract}
\maketitle
\tableofcontents

\section{introduction}

The idea is inspired by the quantum trajectory description of
decoherence\cite{trajectory}. The trajectory was first proposed by Bohm when
he made a suggested interpretation of the quantum theory for hidden
variables\cite{Bohm}. The theory is known as the de Broglie-Bohm(BB)
interpretation of quantum mechanics. In the theory, all particles have
well-defined trajectories. The motions of the particles are governed by the
wave functions that satisfy the Schrodinger equation. Therefore the BB quantum
theory of motion is a suitable tool with which to study coherence and
decoherence\cite{BBdecoherence,BBdecoherence1}. The one reason of the
decoherence is the open quantum system interacts with the environment.
Unfortunately, it is very difficult to deal with decoherence problem using
this quantum trajectory approach because the environment usually involves
large number of degrees of the freedom. To overcome this drawback, we assume
the environment to be the Markovian environment and describe the whole system
by a Markovian master equation. This equation introduces two contributions:
the time-evolution of the coherent state and the quenching factor leading to
decoherence. The quenching factor accounts for physical properties of the
environment and its interaction with the coherent system. Combining the
trajectory theory with reduced density matrix theory yields a new trajectory
called reduced quantum trajectory\cite{trajectory}. The advantage of this
reduced quantum trajectory is that the environment effects are described by a
time-dependent damping factor when these trajectories are applied to the study
of an open quantum system. The reduced quantum trajectory then describes in
detail the evolution of the coherent state. These provide insight in
understanding decoherence.

Recently, the discovery of the high critical temperature of $MgB_{2}$ has
inspired a widely interest in the charged two-condensate
superconductors\cite{two-gap1,two-gap2,TG-E}. The two charged condensates in
the superconductor are tightly bound fermion pairs, or some other charged
bosonic fields such as electronic or protonic Cooper pairs in metallic
hydrogen under certain condition\cite{liquidm}. The charged two-condensate
wave functions correspond to the order parameters of the two different parts
of the Fermi surface. They are coupled because of their electromagnetic
interaction. The system is described by the Ginzberg-Landau model with two
flavors of Cooper pairs\cite{TGBB,TGBB1,TGBB2}. In\cite{TGBB}, the authors
show the charged -condensate Ginzberg-Landau model can be mapped onto a
version of the nonlinear $O\left(  3\right)  $ $\sigma$-model and found this
system possesses a hidden $O\left(  3\right)  $ symmetry. There is a stable
knot solution in the superconductor. This provides us with a new way to
investigate the coherent quantum system.

The topology and geometry play an important role in physics and mathematics
and a great deal of works have been done in the topology and
geometry\cite{Debrus,Eguchi,Morandi,Nakahara,Nash,Schwarz}. Especially, the
vorticity of the vortex in condensate meter and topology of the physical
system have been studied by applying the $\phi-$mapping topological current
theory\cite{Duan,London e,Shi,Shi1}. In this paper, we present the relation
between the current of coherent state and the supercurrent of the two-gap
condensate system. The paper is organized as follows: in part II, the $\phi
-$mapping topological current theory in reduced density trajectory is given.
The current of the coherent state is also presented. In part III, the new
expression of the current is derived. We find this current is similar to the
supercurrent of the charged two-condensate system. In part IV, the symmetry
and the topological properties of the current of the coherent state are
studied based on Faddeev's $O\left(  3\right)  $ nonlinear $\sigma$-model.
Finally, we make a conclusion.

\section{$\phi-$ mapping topological current theory in reduced density
trajectory and the current of the coherent state}

We give a brief review of the reduced quantum trajectory approach as presented
in\cite{trajectory}. We start with the calculation of the reduced density
matrix. The total density matrix of the system is given by%
\begin{equation}
\widehat{\rho}=|\psi_{t}\rangle\times\langle_{t}\psi|, \label{1}%
\end{equation}
where the subscript $t$ denotes the time-dependence of the wave function. We
take the environment degrees of freedom to be $\mathbf{r}_{i}$ $\left(
i=1,\cdots,N\right)  $. The system's reduced density matrix is then given by
tracing the total density matrix $\widehat{\rho}$ over the environment degrees
of freedom, resulting in
\begin{equation}
\widetilde{\rho_{t}}\left(  r,r^{\prime}\right)  =\int\langle\mathbf{r,r}%
_{1}\mathbf{,\cdots r}_{N}|\psi_{t}\rangle\times\langle_{t}\psi|\mathbf{r}%
^{\prime}\mathbf{,r}_{1}^{\prime}\mathbf{,\cdots r}_{N}^{\prime}\rangle
d\mathbf{r}_{1}\cdots d\mathbf{r}_{N}.
\end{equation}
Next the system reduced quantum density current can be derived as follows:%
\begin{equation}
\widetilde{\mathbf{J}_{t}}=\frac{\hbar}{m}\operatorname{Im}\left[  \nabla
_{r}\widetilde{\rho_{t}}\left(  \mathbf{r,r}^{\prime}\right)  \right]
_{r=r^{\prime}}, \label{J}%
\end{equation}
where $\widetilde{\mathbf{J}_{t}}$ satisfies the continuity equation, which is
given as%
\begin{equation}
\partial_{t}\widetilde{\rho_{t}}+\nabla\widetilde{\mathbf{J}_{t}}=0. \label{P}%
\end{equation}
Where $\widetilde{\rho_{t}}$ is the diagonal element of the reduced density
matrix, which provides the measured intensity. We now define the Bohmian-like
velocity using (\ref{J}) and (\ref{P})%
\begin{equation}
\mathbf{V}=\frac{\widetilde{\mathbf{J}_{t}}}{\widetilde{\rho_{t}}}. \label{V}%
\end{equation}
Therefore, we can define a new trajectory associated with the reduced density
matrix%
\begin{equation}
\mathbf{V}=\frac{\hbar}{m}\frac{\operatorname{Im}\left[  \nabla_{r}%
\widetilde{\rho_{t}}\left(  r,r^{\prime}\right)  \right]  }{\operatorname{Re}%
\left[  \widetilde{\rho_{t}}\left(  r,r^{\prime}\right)  \right]
}_{r=r^{\prime}}. \label{V0}%
\end{equation}
which is called reduced quantum trajectory. The disadvantage of this
definition of velocity is it is difficult to give the detailed information at
$\widetilde{\rho_{t}}=0$, or at the zero points of the wave functions. These
zero points are the singularity of the velocity. Next, we will illustrate the
exact expression of the velocity field and its topology at zero point of wave
functions based on $\phi-$mapping topological current theory. To do this, we
must consider the BB quantum mechanics ansatz of the wave function%
\begin{equation}
\left\langle r\right\vert \psi_{t}\rangle=R_{t}\left(  r\right)
e^{iS_{t}\left(  r\right)  /\hbar},
\end{equation}
from the topological viewpoint, the wave function $\left\langle r\right\vert
\psi_{t}\rangle$ is the section of the complex linear bundle, i.e. a section
of 2-dimensional real vector bundle. We can then write this ansatz as%
\begin{equation}
\left\langle r\right\vert \psi_{t}\rangle=\phi^{1}+i\phi^{2}.
\end{equation}
Defining the unit vector of this ansatz yields%
\begin{equation}
n^{1}=\frac{\phi^{1}}{\left\Vert \left\langle r\right\vert \psi_{t}%
\rangle\right\Vert }\text{ \ }n^{2}=\frac{\phi^{2}}{\left\Vert \left\langle
r\right\vert \psi_{t}\rangle\right\Vert }. \label{unit}%
\end{equation}
It is obvious that the unit vector satisfies the condition%
\begin{equation}
n^{a}n^{a}=1\text{ \ \ }a=1,2.
\end{equation}
Using this unit vector and (\ref{V0}), we write the velocity as%
\begin{equation}
\mathbf{V}_{i}=\frac{\hbar}{m}\epsilon_{ab}n^{a}\partial_{i}n^{b}.
\end{equation}
In traditional quantum mechanics, the curl of the velocity vanishes at zero
points of the wave functions. However, the curl of the velocity must be
modified along trajectories because $\boldsymbol{\nabla}\times\mathbf{V}$ need
not vanish at nodal points of the wave function\cite{Shi}. The curl of the
velocity is%
\begin{equation}
\nabla\times\mathbf{V}=\frac{\hbar}{m}\left(  \epsilon^{ijk}\epsilon
_{ab}\partial_{j}n^{a}\partial_{k}n^{b}\right)  \mathbf{e}_{i}. \label{17}%
\end{equation}
Using Eqs.(\ref{unit}), the curl of the velocity can further be written as%
\begin{equation}
\nabla\times\mathbf{V}=\frac{\hbar}{m}\mathbf{e}_{i}\epsilon^{ijk}%
\epsilon_{ab}\frac{\partial}{\partial\phi^{c}}\frac{\partial}{\partial\phi
^{a}}\left(  \ln\left\Vert \phi\right\Vert \right)  \partial_{j}\phi
^{c}\partial_{k}\phi^{b}. \label{18}%
\end{equation}
Defining the vector Jacobian of $\mathbf{\phi}$ by%
\begin{equation}
\mathbf{e}_{i}\epsilon^{ijk}\partial_{j}\phi^{c}\partial_{k}\phi^{b}%
=\epsilon^{cb}\mathbf{D}\left(  \frac{\phi}{x}\right)  , \label{19}%
\end{equation}
and using the well-known result from the Green's function theory in $\phi
-$space, we find that%
\begin{equation}
\frac{\partial}{\partial\phi^{a}}\frac{\partial}{\partial\phi^{a}}%
\ln\left\Vert \phi\right\Vert =2\pi\delta^{2}\left(  \phi\right)  . \label{20}%
\end{equation}
Finally, the curl of the velocity is%
\begin{equation}
\nabla\times\mathbf{V}=\frac{\hbar}{m}2\pi\delta^{2}\left(  \phi\right)
\mathbf{D}\left(  \frac{\phi}{x}\right)  .
\end{equation}
where $\mathbf{D}\left(  \frac{\phi}{x}\right)  $ is the vector Jacobian of
$\mathbf{\phi}$ and satisfies $\epsilon^{ijk}\partial_{j}\phi^{c}\partial
_{k}\phi^{b}=\epsilon^{cb}D^{i}\left(  \frac{\phi}{x}\right)  $. From this, we
learn that the trajectory is at the zero point of the wave function. We
consider, in general, a vector field $\mathbf{\phi}$ on the smooth manifold
$\Sigma$; a zero point $p$ is a singular point of $\mathbf{\phi}$ if
$\mathbf{\phi}_{p}=0$. Consider a closed curve $\gamma\in\Sigma$ encircling
but never touching $p$. In completing one turn along $\gamma$, the vector
field $\mathbf{\phi}$ will turn around itself a certain number of times. By
appropriately assigning signs to the direction of the turn, the algebraic sum
of turns is called index of the curve. It is well known the sum of all the
indices of a chosen vector field $\mathbf{\phi}$ on a compact differentiable
manifold $\Sigma$ equals the Euler-Poincare characteristic of $\Sigma$ that
describes the topological properties of singular points. In application here,
all nodal points form the zero-line of wave function and the zero-line of wave
function is just the locations of trajectories in de Broglie-Bohm quantum
mechanics. The zero points can be denoted by $z_{l}^{i},$ where $l$ represent
the $\ell$ isolated zero points on $\Sigma.$ We assume that $u=(u_{1},u_{2})$
are the coordinates, so that $\delta^{2}\left(  \phi\right)  $ can be expanded
at the zero point%
\begin{equation}
\delta^{2}\left(  \phi\right)  =%
{\displaystyle\sum\limits_{l=1}^{\ell}}
C_{l}\delta^{2}\left(  x^{i}-z_{l}^{i}\right)  , \label{25}%
\end{equation}
where $C_{l}$ are positive coefficients. The winding number of the $lth$
trajectory is%
\begin{align}
W\left(  \phi,z_{i}\right)   &  =C_{l}\int_{\Sigma}\delta^{2}\left(
x^{i}-z_{l}^{i}\right)  D\left(  \frac{\phi}{x}\right)  d^{2}x\label{26}\\
&  =C_{l}D\left(  \frac{\phi}{u}\right)  _{z_{l}}.\nonumber
\end{align}
Here, $D\left(  \frac{\phi}{u}\right)  $ is%
\begin{equation}
D\left(  \frac{\phi}{u}\right)  =\frac{1}{2}\epsilon^{jk}\epsilon_{ab}%
\frac{\partial}{\partial u^{j}}\phi^{a}\frac{\partial}{\partial u^{k}}\phi
^{b}. \label{27}%
\end{equation}
If we let%
\begin{equation}
\left\vert W_{l}\right\vert =\left\vert W\left(  \phi,z_{l}\right)
\right\vert =\beta_{l}, \label{30}%
\end{equation}
where $\beta_{l}$ is Hopf index of $\phi-$mapping on $\Sigma$, with the
interpretation that the function $\mathbf{\phi}$ covers the corresponding
region in $\phi-$space $\beta_{l}$ times when a point covers the neighborhood
of the zero point $z_{l}^{i}$ once. Furthermore, $\delta^{2}\left(
\phi\right)  $ can be expressed as%
\begin{equation}
\delta^{2}\left(  \phi\right)  =%
{\displaystyle\sum\limits_{l=1}^{\ell}}
\frac{\beta_{l}}{\left\vert D\left(  \frac{\phi}{u}\right)  \right\vert
_{z_{l}}}\delta^{2}\left(  x^{i}-z_{l}^{i}\right)  . \label{31}%
\end{equation}
Let us define%
\begin{equation}
\eta_{l}=signD\left(  \frac{\phi}{u}\right)  _{z_{l}}=\frac{D\left(
\frac{\phi}{u}\right)  }{\left\vert D\left(  \frac{\phi}{u}\right)
\right\vert }_{z_{l}}=\pm1, \label{32}%
\end{equation}
which is called the Brouwer degree of the map $x\rightarrow\phi\left(
x\right)  $. Finally, the vorticity of the velocity at the zero points on
$\Sigma$\ is%
\begin{equation}
\Gamma=\int_{\Sigma}\left(  \nabla\times\mathbf{V}\right)  \cdot
d\mathbf{S}=\frac{h}{m}%
{\displaystyle\sum\limits_{l}}
\beta_{l}\eta_{l}=\frac{h}{m}W, \label{W}%
\end{equation}
where $W$ is the winding number of the zero points of the trajectories on
$\Sigma$. The zero points on the plane can be seen as the topological
solutions of the equation $\delta^{2}\left(  \phi\right)  $ and can be written
as%
\begin{align}
\phi^{1}\left(  x^{\mu}\right)   &  =0,\nonumber\\
\phi^{2}\left(  x^{\mu}\right)   &  =0, \label{E}%
\end{align}
where $\mu=1,2,3..$

Considering a quantum system in the double-slit experiment, the system is
described by the coherent state of a particle and the state of the
environment. The coherent state of a particle is%
\begin{equation}
|\Psi_{t}\rangle=c_{1}|\psi_{1,t}\rangle+c_{2}|\psi_{2,t}\rangle,
\end{equation}
where the coefficients $c_{a}$ satisfies the condition%
\begin{equation}
\left\vert c_{1}\right\vert ^{2}+\left\vert c_{2}\right\vert ^{2}%
=1.\label{inerproduct}%
\end{equation}
We assume the environment states are subject to the elastic system-environment
scattering conditions\cite{trajectory}, then only the environment state will
evolve with time. The environment state associated with each partial wave is
denoted by $|H_{\alpha}\rangle$. The initial state of the environment states
can be given by%
\begin{equation}
|H_{1}\rangle=|H_{2}\rangle=|H_{0}\rangle.
\end{equation}
Using BB quantum mechanics anzatz, the coherent state can be described without
considering the interaction between coherence states and the environment
\begin{equation}
\Psi_{t}\left(  r\right)  =\langle r|\Psi_{t}\rangle.
\end{equation}
The density matrix associated with coherent state is
\begin{equation}
\rho_{t}\left(  \mathbf{r,r}^{\prime}\right)  =\Psi_{t}\left(  \mathbf{r,r}%
^{\prime}\right)  \left[  \Psi_{t}\left(  \mathbf{r,r}^{\prime}\right)
\right]  ^{\ast}.\label{density}%
\end{equation}
The diagonal element of this density matrix is the measured intensity. We
write it as
\begin{equation}
\rho_{t}=\left\vert c_{1}\right\vert ^{2}\left\vert \psi_{1,t}\right\vert
^{2}+\left\vert c_{2}\right\vert ^{2}\left\vert \psi_{2,t}\right\vert
^{2}+2\left\vert c_{1}\right\vert \left\vert c_{2}\right\vert \left\vert
\psi_{1,t}\right\vert _{t}\left\vert \psi_{2,t}\right\vert \cos\delta
_{t},\label{D1}%
\end{equation}
where $\delta_{t}$ is the time-dependent phase shift between the
partial\ waves. Similarly the partial wave function $\psi_{i,t}$ can be
written as%
\begin{equation}
\psi_{i,t}=\langle r|\psi_{i,t}\rangle.
\end{equation}
In addition to writing the measured intensity for $\Psi_{t}\left(  r\right)
$, we define the measured intensity of the partial wave function $\rho
_{t}^{\left(  i\right)  }$ as
\begin{equation}
\rho_{t}^{\left(  i\right)  }=\psi_{i,t}^{\ast}\psi_{i,t}\text{ \ }i=1,2,.
\end{equation}
The partial wave function can also be expressed as%
\begin{equation}
\psi_{i,t}=\phi_{i,t}^{1}+i\phi_{i,t}^{2}.
\end{equation}
Recalling (\ref{unit}), the unit vector $\mathbf{n}_{\left(  i\right)  }$ of
the partial wave function $\psi_{i,t}$ is defined by%
\begin{equation}
n_{\left(  i\right)  }^{1}=\frac{\phi_{i,t}^{1}}{\left\Vert \psi
_{i,t}\right\Vert },n_{\left(  i\right)  }^{2}=\frac{\phi_{i,t}^{2}%
}{\left\Vert \psi_{i,t}\right\Vert }.\label{unit1}%
\end{equation}

The general initial coherent states get entangled with the environment states
when the environment is considered. The initial entangled state is%
\begin{equation}
|\Psi\rangle=|\Psi_{0}\rangle\otimes|H_{0}\rangle,
\end{equation}
where $|\Psi_{0}\rangle$ is the wave function $|\Psi_{t}\rangle$ at time
$t=0$. The time-dependence of the entangled state is%
\begin{equation}
|\Psi_{t}\rangle=c_{1}|\psi_{1,t}\rangle\otimes|H_{1,t}\rangle+c_{2}%
|\psi_{2,t}\rangle\otimes|H_{2,t}\rangle, \label{density1}%
\end{equation}
where $|H_{i,t}\rangle$ is the time-dependent environment. Then we obtain the
measured intensity of the entangled state by tracing the full density matrix
over the environment state%
\begin{equation}
\widetilde{\rho_{t}}=%
{\displaystyle\sum\limits_{a=1}^{2}}
\langle H_{a,t}\left\vert \widehat{\rho}\right\vert H_{a,t}\rangle.
\label{reduced density}%
\end{equation}
Substitute (\ref{density1}) and (\ref{1}) into (\ref{reduced density}), one
obtains the measured intensity by tracing the total density matrix over the
environmental degrees of freedom%
\begin{equation}
\widetilde{\rho}_{t}=\left(  1+\left\vert a_{t}\right\vert ^{2}\right)
{\displaystyle\sum\limits_{i=1}^{2}}
\left\vert c_{i}\right\vert ^{2}\psi_{i,t}^{\ast}\psi_{i,t}+2a_{t}c_{1}%
c_{2}^{\ast}\psi_{1,t}\psi_{2,t}^{\ast}+c.c.. \label{density2}%
\end{equation}
This equation means the interaction between the coherence state and the
environment is the reason of the decoherence. The coefficient $a_{t}=\langle
H_{2,t}|H_{1,t}\rangle$ is called the damping factor and indicates the degree
of coherence. The cross terms $c_{1}c_{2}^{\ast}\psi_{1,t}\psi_{2,t}^{\ast}$
and its conjugate complex in (\ref{density2}) disappear, when $a_{t}=0,$ the
coherent state suffers a total loss of coherence. If one introduces the
coherence time $\tau,$ then this damping factor can be written as
$a_{t}=e^{-t/\tau}$. By using (\ref{V0}) and (\ref{density2}), the current is
given by%
\begin{align}
\mathbf{J}  &  \mathbf{=}\widetilde{\rho_{t}}\mathbf{V}=\frac{i\left(
1+\left\vert a_{t}\right\vert ^{2}\right)  \hbar}{2m}%
{\displaystyle\sum\limits_{i=1}^{2}}
\left\vert c_{i}\right\vert ^{2}\left[  \left(  \psi_{i,t}^{\ast}\nabla
\psi_{i,t}-\psi_{i,t}\nabla\psi_{i,t}^{\ast}\right)  \right] \label{29}\\
&  +\frac{i\hbar}{m}\left\vert a_{t}\right\vert c_{1}c_{2}^{\ast}\left[
\left(  \psi_{2,t}^{\ast}\nabla\psi_{1,t}-\psi_{1,t}\nabla\psi_{2,t}^{\ast
}\right)  \right]  +C.C..\nonumber
\end{align}

\section{\bigskip the current as a supercurrent in two-condensate system}

From equation (\ref{29}), the current is seen to be expressed as a sum of two
contributions: the first term, which does not include the cross term of the
partial wave functions, will be denoted by $\mathbf{J}_{1}$
\begin{equation}
\mathbf{J}_{1}=\frac{i\left(  1+\left\vert a_{t}\right\vert ^{2}\right)
\hbar}{2m}%
{\displaystyle\sum\limits_{i=1}^{2}}
\left\vert c_{i}\right\vert ^{2}\left[  \left(  \psi_{i,t}^{\ast}\nabla
\psi_{i,t}-\psi_{i,t}\nabla\psi_{i,t}^{\ast}\right)  \right]  ,
\end{equation}
and the second term, which includes the cross term which indicates the
coherent effects, will be written by $\mathbf{J}_{2}$%
\begin{align}
\mathbf{J}_{2} &  =\frac{i\hbar}{m}\left\vert a_{t}\right\vert c_{1}%
c_{2}^{\ast}\left[  \left(  \psi_{2,t}^{\ast}\nabla\psi_{1,t}-\psi_{1,t}%
\nabla\psi_{2,t}^{\ast}\right)  \right]  \nonumber\\
&  +\frac{i\hbar}{m}\left\vert a_{t}\right\vert c_{1}^{\ast}c_{2}\left[
\left(  \psi_{1,t}^{\ast}\nabla\psi_{2,t}-\psi_{2,t}\nabla\psi_{1,t}^{\ast
}\right)  \right]  .
\end{align}
In terms of the partial measured intensity of the partial wave function
$\rho_{t}^{\left(  i\right)  }$, $\mathbf{J}_{1}$ is
\begin{equation}
\mathbf{J}_{1}=\frac{i\left(  1+\left\vert a_{t}\right\vert ^{2}\right)
\hbar}{2m}\left[  \left\vert c_{1}\right\vert ^{2}\left(  \psi_{1,t}^{\ast
}\psi_{1,t}\right)  \frac{\left(  \psi_{1,t}^{\ast}\nabla\psi_{1,t}-\psi
_{1,t}\nabla\psi_{1,t}^{\ast}\right)  }{\left(  \psi_{1,t}^{\ast}\psi
_{1,t}\right)  }+\left\vert c_{2}\right\vert ^{2}\left(  \psi_{2,t}^{\ast}%
\psi_{2,t}\right)  \frac{\left(  \psi_{2,t}^{\ast}\nabla\psi_{2,t}-\psi
_{2,t}\nabla\psi_{2,t}^{\ast}\right)  }{\left(  \psi_{2,t}^{\ast}\psi
_{2,t}\right)  }\right]  .\label{V1}%
\end{equation}
In a similar manner, $\mathbf{J}_{2\text{ }}$ is also rewritten as%
\begin{align}
\mathbf{J}_{2} &  =\frac{i\hbar}{m}\left\vert a_{t}\right\vert \left[  \left(
c_{1}c_{2}^{\ast}\psi_{1,t}\psi_{2,t}^{\ast}\frac{\psi_{1,t}^{\ast}\nabla
\psi_{1,t}}{\psi_{1,t}^{\ast}\psi_{1,t}}-c_{1}^{\ast}c_{2}\psi_{1,t}^{\ast
}\psi_{2,t}\frac{\psi_{1,t}\nabla\psi_{1,t}^{\ast}}{\psi_{1,t}^{\ast}%
\psi_{1,t}}\right)  \right]  \nonumber\\
&  +\frac{i\hbar}{m}\left\vert a_{t}\right\vert \left[  \left(  c_{1}^{\ast
}c_{2}\psi_{2,t}\psi_{1,t}^{\ast}\frac{\psi_{2,t}^{\ast}\nabla\psi_{2,t}}%
{\psi_{2,t}^{\ast}\psi_{2,t}}-c_{1}c_{2}^{\ast}\psi_{2,t}^{\ast}\psi
_{1,t}\frac{\psi_{2,t}\nabla\psi_{2,t}^{\ast}}{\psi_{2,t}^{\ast}\psi_{2,t}%
}\right)  \right]  .
\end{align}
Let us define the complex variable $\Lambda=c_{1}c_{2}^{\ast}\psi_{1,t}%
\psi_{2,t}^{\ast};$ then $\Lambda^{\ast}=c_{1}^{\ast}c_{2}\psi_{1,t}^{\ast
}\psi_{2,t}$, the current $\mathbf{J}_{2}$ can be rewritten as
\begin{align}
\mathbf{J}_{2} &  =\frac{i\hbar}{m}\left\vert a_{t}\right\vert \left[  \left(
\Lambda\frac{\psi_{1,t}^{\ast}\nabla\psi_{1,t}}{\psi_{1,t}^{\ast}\psi_{1,t}%
}-\Lambda^{\ast}\frac{\psi_{1,t}\nabla\psi_{1,t}^{\ast}}{\psi_{1,t}^{\ast}%
\psi_{1,t}}\right)  \right]  \nonumber\\
&  +\frac{i\hbar}{m}\left\vert a_{t}\right\vert \left[  \left(  \Lambda^{\ast
}\frac{\psi_{2,t}^{\ast}\nabla\psi_{2,t}}{\psi_{2,t}^{\ast}\psi_{2,t}}%
-\Lambda\frac{\psi_{2,t}\nabla\psi_{2,t}^{\ast}}{\psi_{2,t}^{\ast}\psi_{2,t}%
}\right)  \right]  .\label{velocity}%
\end{align}
It is convenient to write $\Lambda=\Lambda_{1}+i\Lambda_{2}$ and
$\Lambda^{\ast}=\Lambda_{1}-i\Lambda_{2}$, where $\Lambda_{1}$ and
$\Lambda_{2}$ are real numbers. Substituting $\Lambda_{1}$ and $\Lambda_{2}$
into (\ref{velocity}), one obtains%
\begin{align}
\mathbf{J}_{2} &  =\frac{i\hbar}{m}\left\vert a_{t}\right\vert \left[
\Lambda_{1}\left(  \frac{\psi_{1,t}^{\ast}\nabla\psi_{1,t}}{\psi_{1,t}^{\ast
}\psi_{1,t}}-\frac{\psi_{1,t}\nabla\psi_{1,t}^{\ast}}{\psi_{1,t}^{\ast}%
\psi_{1,t}}\right)  +i\Lambda_{2}\left(  \frac{\psi_{1,t}^{\ast}\nabla
\psi_{1,t}}{\psi_{1,t}^{\ast}\psi_{1,t}}+\frac{\psi_{1,t}\nabla\psi
_{1,t}^{\ast}}{\psi_{1,t}^{\ast}\psi_{1,t}}\right)  \right]  \nonumber\\
&  +\frac{i\hbar}{m}\left\vert a_{t}\right\vert \left[  \Lambda_{1}\left(
\frac{\psi_{2,t}^{\ast}\nabla\psi_{2,t}}{\psi_{2,t}^{\ast}\psi_{2,t}}%
-\frac{\psi_{2,t}\nabla\psi_{2,t}^{\ast}}{\psi_{2,t}^{\ast}\psi_{2,t}}\right)
-i\Lambda_{2}\left(  \frac{\psi_{2,t}^{\ast}\nabla\psi_{2,t}}{\psi_{2,t}%
^{\ast}\psi_{2,t}}+\frac{\psi_{2,t}\nabla\psi_{2,t}^{\ast}}{\psi_{2,t}^{\ast
}\psi_{2,t}}\right)  \right]  .
\end{align}
In term of the relations%
\begin{equation}
\nabla\ln\left(  \psi_{i,t}^{\ast}\psi_{i,t}\right)  =\left(  \frac{\psi
_{i,t}^{\ast}\nabla\psi_{i,t}}{\psi_{i,t}^{\ast}\psi_{i,t}}+\frac{\psi
_{i,t}\nabla\psi_{i,t}^{\ast}}{\psi_{i,t}^{\ast}\psi_{i,t}}\right)  \text{
\ }i=1,2,
\end{equation}
finally, $\mathbf{J}_{2\text{ }}$can be expressed by%
\begin{align}
\mathbf{J}_{2} &  =\frac{i\hbar}{m}\left\vert a_{t}\right\vert \Lambda
_{1}\left[  \left(  \frac{\psi_{1,t}^{\ast}\nabla\psi_{1,t}}{\psi_{1,t}^{\ast
}\psi_{1,t}}-\frac{\psi_{1,t}\nabla\psi_{1,t}^{\ast}}{\psi_{1,t}^{\ast}%
\psi_{1,t}}\right)  +\left(  \frac{\psi_{2,t}^{\ast}\nabla\psi_{2,t}}%
{\psi_{2,t}^{\ast}\psi_{2,t}}-\frac{\psi_{2,t}\nabla\psi_{2,t}^{\ast}}%
{\psi_{2,t}^{\ast}\psi_{2,t}}\right)  \right]  \nonumber\\
&  +\frac{\hbar}{m}\left\vert a_{t}\right\vert \Lambda_{2}\mathbf{\nabla
}\left[  \ln\left(  \frac{\psi_{1,t}^{\ast}\psi_{1,t}}{\psi_{2,t}^{\ast}%
\psi_{2,t}}\right)  \right]  .\label{D2}%
\end{align}
This formula shows there is a topological reason leading to the decoherence.
The new parameter $\Lambda_{1}$ can be used to indicate the coherent degree.
This parameter also can be called damping factor, but it is very different
from the parameter $a_{t}$. The parameter $a_{t}$ relates to the degrees of
the freedom of the environment. But from (\ref{D1}), the parameter
$\Lambda_{1}$ relates to the phase shift of the partial wave functions. The
parameter $\Lambda_{1}$ is indispensable to give the exact expression
(\ref{D2}), which is essential for giving the topological structure of the
current. Then the parameter $\Lambda_{1}$ is important to the topological
structure of the current, but the parameter $a_{t}$ has nothing to do with the
topological structure. In addition, we find $\hbar\mathbf{\nabla}\left[
\ln\left(  \frac{\psi_{1,t}^{\ast}\psi_{1,t}}{\psi_{2,t}^{\ast}\psi_{2,t}%
}\right)  \right]  $ is a vector, then a new $U\left(  1\right)  $ gauge
potential is defined by
\begin{equation}
\mathbf{A}=\hbar\mathbf{\nabla}\left[  \ln\left(  \frac{\psi_{1,t}^{\ast}%
\psi_{1,t}}{\psi_{2,t}^{\ast}\psi_{2,t}}\right)  \right]  .
\end{equation}
Therefore, the current $\mathbf{J}_{2}$ is
\begin{align}
\mathbf{J}_{2} &  =\frac{i\hbar}{m}\left\vert a_{t}\right\vert \Lambda
_{1}\left[  \left(  \frac{\psi_{1,t}^{\ast}\nabla\psi_{1,t}}{\psi_{1,t}^{\ast
}\psi_{1,t}}-\frac{\psi_{1,t}\nabla\psi_{1,t}^{\ast}}{\psi_{1,t}^{\ast}%
\psi_{1,t}}\right)  +\left(  \frac{\psi_{2,t}^{\ast}\nabla\psi_{2,t}}%
{\psi_{2,t}^{\ast}\psi_{2,t}}-\frac{\psi_{2,t}\nabla\psi_{2,t}^{\ast}}%
{\psi_{2,t}^{\ast}\psi_{2,t}}\right)  \right]  \nonumber\\
&  +\frac{1}{m}\left\vert a_{t}\right\vert \Lambda_{2}\mathbf{A.}%
\end{align}
We assume the system is in the coherence, that is to say, the damping factor
$\left\vert a_{t}\right\vert =1.$ The current using the measured intensity of
the partial wave function $\rho_{t}^{\left(  i\right)  }$ is%
\begin{align}
\mathbf{J} &  =\frac{i\hbar}{m}\left(  \left\vert c_{1}\right\vert ^{2}%
\rho_{t}^{\left(  1\right)  }+\Lambda_{1}\right)  \left[  \frac{\left(
\psi_{1,t}^{\ast}\nabla\psi_{1,t}-\psi_{1,t}\nabla\psi_{1,t}^{\ast}\right)
}{\psi_{1,t}^{\ast}\psi_{1,t}}\right]  +\frac{i\hbar}{m}\left(  \left\vert
c_{2}\right\vert ^{2}\rho_{t}^{\left(  2\right)  }+\Lambda_{1}\right)  \left[
\frac{\left(  \psi_{2,t}^{\ast}\nabla\psi_{2,t}-\psi_{2,t}\nabla\psi
_{2,t}^{\ast}\right)  }{\psi_{2,t}^{\ast}\psi_{2,t}}\right]  \nonumber\\
&  +\frac{1}{m}\Lambda_{2}\mathbf{A}%
\end{align}
In order to study the current in detail, we define new charges $q_{1}%
=\frac{\left(  \left\vert c_{1}\right\vert ^{2}\rho_{t}^{\left(  1\right)
}+\Lambda_{1}\right)  }{\rho_{t}^{\left(  1\right)  }}$ and $q_{2}%
=\frac{\left(  \left\vert c_{2}\right\vert ^{2}\rho_{t}^{\left(  2\right)
}+\Lambda_{1}\right)  }{\rho_{t}^{\left(  2\right)  }}.$ Then the new
$U\left(  1\right)  $ gauge potential is given by%
\[
\widetilde{\mathbf{A}}=\frac{\Lambda_{2}}{4\left(  q_{1}^{2}+q_{2}^{2}\right)
\left(  \left\vert \psi_{1,t}\right\vert ^{2}+\left\vert \psi_{2,t}\right\vert
^{2}\right)  }\mathbf{A.}%
\]
Based on the new gauge potential $\widetilde{\mathbf{A}}\mathbf{,}$ The
current of the coherent system can be expressed by
\begin{align}
\mathbf{J} &  =\frac{i\hbar q_{1}}{m}\left(  \psi_{1,t}^{\ast}\nabla\psi
_{1,t}-\psi_{1,t}\nabla\psi_{1,t}^{\ast}\right)  +\frac{i\hbar q_{2}}%
{m}\left(  \psi_{2,t}^{\ast}\nabla\psi_{2,t}-\psi_{2,t}\nabla\psi_{2,t}^{\ast
}\right)  \nonumber\\
&  +\frac{4\left(  q_{1}^{2}+q_{2}^{2}\right)  }{m}\left(  \left\vert
\psi_{1,t}\right\vert ^{2}+\left\vert \psi_{2,t}\right\vert ^{2}\right)
\widetilde{\mathbf{A}}.\label{totalcurrent}%
\end{align}
However, we find the total current can be deduced from the following free
energy%
\begin{equation}
F=\left[  \frac{1}{2m}\left\vert \left(  \hbar\partial_{k}+i\frac{2q_{1}}%
{c}\widetilde{\mathbf{A}}_{k}\right)  \psi_{1,t}\right\vert ^{2}+\frac{1}%
{2m}\left\vert \left(  \hbar\partial_{k}+i\frac{2q_{2}}{c}\widetilde
{\mathbf{A}}_{k}\right)  \psi_{2,t}\right\vert ^{2}+V\left(  \left\vert
\psi_{a,t}\right\vert ^{2}\right)  +\frac{\widetilde{\mathbf{B}}^{2}}{8\pi
}\right]  ,\label{free energy}%
\end{equation}
where $\widetilde{\mathbf{B}}=\mathbf{\nabla\times}\widetilde{\mathbf{A}}$ is
$U(1)$ gauge field. The potential $V\left(  \left\vert \psi_{a,t}\right\vert
^{2}\right)  $ is%
\begin{equation}
V\left(  \left\vert \psi_{a,t}\right\vert ^{2}\right)  =-b_{a}\left\vert
\psi_{a,t}\right\vert ^{2}+\frac{c_{a}}{2}\left\vert \psi_{a,t}\right\vert
^{4}.\text{ \ }a=1,2.
\end{equation}
It is well known this free energy is called Ginzberg-Landau free energy, which
is used to describe the charged two-condensate Bose system \cite{TGBB}. The
total current (\ref{totalcurrent}) of quantum coherent system is similar to
the supercurrent of the charged two-condensate Bose system. In two-condensate
superconductor, the charged two-condensate wave functions, or charged order
parameters, can carry the electronic charges. The interaction of charged order
parameters is mediated by the electromagnetic potential $\mathbf{A}_{e}$. In
this description, we find the coherent system interacting with the environment
is similar to the two-condensate superconductor. The partial wave functions
can be seen as the charged order parameters. The partial wave functions are
weakly-coupled because they carry the charges $q_{1}$ and $q_{2}$, which is
different from the electronic charge. The interaction of the partial wave
functions is mediated by the new $U\left(  1\right)  $ gauge potential
$\widetilde{\mathbf{A}}$, not the electromagnetic potential.

\section{the symmetry of the current and its topology}

In this section, we try to study the free energy, symmetry and the topological
properties of the current of the coherent state. Let us define the partial
wave function as%
\begin{equation}
\psi_{a,t}=\sqrt{2m}\rho\xi_{a}\text{ \ }a=1,2,
\end{equation}
where the complex variable $\xi_{a}=\left\vert \xi_{a}\right\vert e^{i\theta}%
$. The modular $\rho$ is%
\begin{equation}
\rho=\frac{1}{2}\left(  \frac{\left\vert \psi_{1,t}\right\vert ^{2}}{m}%
+\frac{\left\vert \psi_{2,t}\right\vert ^{2}}{m}\right)  .
\end{equation}
By using these new variables, the Ginzberg-Landau-like free energy of the
coherent state is given as
\begin{align}
F &  =\hbar^{2}\left(  \partial\rho\right)  ^{2}+\hbar^{2}\rho^{2}\left\vert
\left(  \partial_{k}+i\frac{2q_{1}}{\hbar c}\widetilde{\mathbf{A}}\right)
\xi_{1}\right\vert ^{2}+\hbar^{2}\rho^{2}\left\vert \left(  \partial
_{k}+i\frac{2q_{2}}{\hbar c}\widetilde{\mathbf{A}}\right)  \xi_{2}\right\vert
\nonumber\\
&  +V\left(  \left\vert \psi_{a,t}\right\vert ^{2}\right)  +\frac
{\widetilde{\mathbf{B}}^{2}}{8\pi}.\label{FE}%
\end{align}
It can be rewritten by%
\begin{align}
F &  =\hbar^{2}\left(  \partial\rho\right)  ^{2}+\hbar^{2}\rho^{2}\left(
\left\vert \partial\xi_{1}\right\vert ^{2}+\left\vert \partial\xi
_{2}\right\vert ^{2}\right)  +V\left(  \left\vert \psi_{a,t}\right\vert
^{2}\right)  +\frac{\widetilde{\mathbf{B}}^{2}}{8\pi}\nonumber\\
&  +\hbar^{2}\rho^{2}\left[  i\frac{2q_{1}}{\hbar c}\left(  \widetilde
{\mathbf{A}}\xi_{1}\partial\xi_{1}^{\ast}-\widetilde{\mathbf{A}}\xi_{1}^{\ast
}\partial\xi_{1}\right)  +\frac{4q_{1}^{2}}{\hbar^{2}c^{2}}\left\vert \xi
_{1}\right\vert ^{2}\widetilde{\mathbf{A}}\right]  \nonumber\\
&  +\hbar^{2}\rho^{2}\left[  i\frac{2q_{2}}{\hbar c}\left(  \widetilde
{\mathbf{A}}\xi_{2}\partial\xi_{2}^{\ast}-\widetilde{\mathbf{A}}\xi_{2}^{\ast
}\partial\xi_{2}\right)  +\frac{4q_{2}^{2}}{\hbar^{2}c^{2}}\left\vert \xi
_{2}\right\vert ^{2}\widetilde{\mathbf{A}}\right]  .
\end{align}
The supercurrent of the free energy can be derived as%
\begin{equation}
\mathbf{J=}i\hbar^{2}\rho^{2}\left[  \frac{2q_{1}}{\hbar c}\left(  \xi
_{1}\partial\xi_{1}^{\ast}-\xi_{1}^{\ast}\partial\xi_{1}\right)  +\frac
{2q_{2}}{\hbar c}\left(  \xi_{2}\partial\xi_{2}^{\ast}-\xi_{2}^{\ast}%
\partial\xi_{2}\right)  \right]  +\hbar^{2}\rho^{2}\left(  \frac{4q_{1}^{2}%
}{\hbar^{2}c^{2}}\left\vert \xi_{1}\right\vert ^{2}+\frac{4q_{2}^{2}}%
{\hbar^{2}c^{2}}\left\vert \xi_{2}\right\vert ^{2}\right)  \widetilde
{\mathbf{A}}.
\end{equation}
Let $\Delta=\left(  \frac{4q_{1}^{2}}{\hbar^{2}c^{2}}\left\vert \xi
_{1}\right\vert ^{2}+\frac{4q_{2}^{2}}{\hbar^{2}c^{2}}\left\vert \xi
_{2}\right\vert ^{2}\right)  $, the new supercurrent $\widetilde{\mathbf{J}}$
is given as%
\begin{align}
\widetilde{\mathbf{J}} &  =\frac{\mathbf{J}}{\hbar^{2}\rho^{2}\Delta
}\nonumber\\
&  =i\left[  \frac{2q_{1}}{\hbar c\Delta}\left(  \xi_{1}\partial\xi_{1}^{\ast
}-\xi_{1}^{\ast}\partial\xi_{1}\right)  +\frac{2q_{2}}{\hbar c\Delta}\left(
\xi_{2}\partial\xi_{2}^{\ast}-\xi_{2}^{\ast}\partial\xi_{2}\right)  \right]
+\widetilde{\mathbf{A}}.
\end{align}
To find the symmetry of the coherent system, a new complex variable
$\widetilde{\xi_{a}}$ is defined by
\begin{equation}
\widetilde{\xi_{a}}=\sqrt{\frac{2q_{a}}{\hbar\Delta Qc}}\xi_{a},
\end{equation}
where the real number $Q$ guarantee the new partial wave functions satisfy%
\begin{equation}
\left\vert \widetilde{\xi_{1}}\right\vert ^{2}+\left\vert \widetilde{\xi_{2}%
}\right\vert ^{2}=1.
\end{equation}
In terms of the new complex variable, the supercurrent $\widetilde{\mathbf{J}%
}$ is%
\begin{equation}
\widetilde{\mathbf{J}}\mathbf{=}iQ\left[  \left(  \widetilde{\xi}_{1}%
\partial\widetilde{\xi}_{1}^{\ast}-\widetilde{\xi}_{1}^{\ast}\partial
\widetilde{\xi}_{1}\right)  +\left(  \widetilde{\xi}_{2}\partial\widetilde
{\xi}_{2}^{\ast}-\widetilde{\xi}_{2}^{\ast}\partial\widetilde{\xi}_{2}\right)
\right]  +\widetilde{\mathbf{A}}.
\end{equation}
Next we define a gauge invariant unit vector $\widetilde{\mathbf{n}}$%
\begin{equation}
\widetilde{\mathbf{n}}\mathbf{=}\left(  \overline{\widetilde{\xi}%
},\mathbf{\sigma}\widetilde{\xi}\right)  ,
\end{equation}
where $\overline{\widetilde{\xi}}=\left(  \widetilde{\xi}_{1}^{\ast
},\widetilde{\xi}_{2}^{\ast}\right)  $ and $\sigma$ are the Pauli matrices. It
is obvious the unit vector satisfies%
\[
\widetilde{\mathbf{n}}\bullet\widetilde{\mathbf{n}}=1.
\]
Then a new vector $\mathbf{C}$ can be defined by%
\begin{equation}
\mathbf{C=}Q\frac{\mathbf{j}}{2}+\widetilde{\mathbf{A}},
\end{equation}
where $\mathbf{j=}i\left[  \left(  \widetilde{\xi}_{1}\partial\widetilde{\xi
}_{1}^{\ast}-\widetilde{\xi}_{1}^{\ast}\partial\widetilde{\xi}_{1}\right)
+\left(  \widetilde{\xi}_{2}\partial\widetilde{\xi}_{2}^{\ast}-\widetilde{\xi
}_{2}^{\ast}\partial\widetilde{\xi}_{2}\right)  \right]  .$ We add and
subtract from (\ref{FE}) a term $\frac{1}{4}\hbar^{2}\rho^{2}Q^{2}\Delta
^{2}\mathbf{j}^{2}$, the two charged free energy of the coherent state can be
expressed with these new variables%
\begin{align}
F &  =\hbar^{2}\left(  \partial\rho\right)  ^{2}+\frac{\hbar^{2}\rho^{2}%
Q^{2}\Delta^{2}}{4}\left(  \partial\widetilde{\mathbf{n}}\right)  ^{2}%
+\frac{1}{8\pi}\left[  \left(  \partial_{i}C_{j}-\partial_{j}C_{i}\right)
-\frac{Q}{4}\widetilde{\mathbf{n}}\cdot\mathbf{\partial}_{i}\widetilde
{\mathbf{n}}\times\mathbf{\partial}_{j}\widetilde{\mathbf{n}}\right]
+\hbar^{2}\rho^{2}\Delta\mathbf{C}^{2}+V\nonumber\\
&  +\hbar^{2}\rho^{2}\left[  \left(  1-\frac{2q_{1}}{\hbar c}\right)
\left\vert \partial\xi_{1}\right\vert ^{2}+\left(  1-\frac{2q_{2}}{\hbar
c}\right)  \left\vert \partial\xi_{2}\right\vert ^{2}\right]  .
\end{align}
Considering the London limit, we have $\partial\rho=0$, the free energy is
given by%
\begin{align}
F &  =\frac{\hbar^{2}\rho^{2}Q^{2}\Delta^{2}}{4}\left(  \partial
\widetilde{\mathbf{n}}\right)  ^{2}+\frac{1}{8\pi}\left[  \left(  \partial
_{i}C_{j}-\partial_{j}C_{i}\right)  -\frac{Q}{4}\widetilde{\mathbf{n}}%
\cdot\mathbf{\partial}_{i}\widetilde{\mathbf{n}}\times\mathbf{\partial}%
_{j}\widetilde{\mathbf{n}}\right]  +\hbar^{2}\rho^{2}\Delta\mathbf{C}%
^{2}+V\nonumber\\
&  +\hbar^{2}\rho^{2}\left[  \left(  1-\frac{2q_{1}}{\hbar c}\right)
\left\vert \partial\xi_{1}\right\vert ^{2}+\left(  1-\frac{2q_{2}}{\hbar
c}\right)  \left\vert \partial\xi_{2}\right\vert ^{2}\right]  .
\end{align}
Finally, we find there is a stable knotted solution in coherent system, which
is described by the Skyme-Faddeev-Niemi action%
\begin{equation}
F_{0}=\frac{\rho^{2}\hbar^{2}Q^{2}\Delta^{2}}{4}\left(  \partial
\widetilde{\mathbf{n}}\right)  ^{2}+\frac{Q}{32\pi}\widetilde{\mathbf{n}}%
\cdot\mathbf{\partial}_{i}\widetilde{\mathbf{n}}\times\mathbf{\partial}%
_{j}\widetilde{\mathbf{n}}.
\end{equation}
The knotted solution displays a $O\left(  3\right)  $ symmetry in the free
energy. The knotted solution is just the nontrivial map
\begin{equation}
\widetilde{\mathbf{n}}:S^{3}\rightarrow S^{2}.
\end{equation}
The boundary condition of this knotted solution is%
\begin{equation}
\widetilde{\mathbf{n}}\left(  x\right)  \rightarrow\widetilde{\mathbf{n}}%
_{0}\text{ \ \ }\mathbf{x\rightarrow\infty}%
\end{equation}
where $\mathbf{n}_{0}$ is the constant vector in spatial direction. The
knotted solution has an important relation to the current of the coherent
state. It is convenient to write the current as%
\begin{equation}
\mathbf{J}=\mathbf{J}_{1}+\mathbf{J}_{2},
\end{equation}
where
\begin{equation}
\mathbf{J}_{1}=\frac{\hbar q_{1}\rho_{t}^{\left(  1\right)  }}{im}\left[
\frac{\left(  \psi_{1,t}^{\ast}\nabla\psi_{1,t}-\psi_{1,t}\nabla\psi
_{1,t}^{\ast}\right)  }{\psi_{1,t}^{\ast}\psi_{1,t}}\right]  +\frac{4q_{1}%
^{2}}{m}\left\vert \psi_{1,t}\right\vert ^{2}\widetilde{\mathbf{A}}%
\end{equation}
and%
\begin{equation}
\mathbf{J}_{2}=\frac{\hbar q_{2}\rho_{t}^{\left(  2\right)  }}{im}\left[
\frac{\left(  \psi_{2,t}^{\ast}\nabla\psi_{2,t}-\psi_{2,t}\nabla\psi
_{2,t}^{\ast}\right)  }{\psi_{2,t}^{\ast}\psi_{2,t}}\right]  +\frac{4q_{2}%
^{2}}{m}\left\vert \psi_{2,t}\right\vert ^{2}\widetilde{\mathbf{A}}.
\end{equation}
Recalling the unit vector $\mathbf{n}_{\left(  i\right)  }$, these components
can be derived as%
\begin{equation}
\mathbf{J}_{1}=\frac{\hbar q_{1}\rho_{t}^{\left(  1\right)  }}{m}\epsilon
_{ab}n_{(1)}^{a}\partial_{i}n_{(1)}^{b}+\frac{4q_{1}^{2}}{m}\left\vert
\psi_{1,t}\right\vert ^{2}\widetilde{\mathbf{A}}%
\end{equation}
and%
\begin{equation}
\mathbf{J}_{2}=\frac{\hbar q_{2}\rho_{t}^{\left(  2\right)  }}{m}\epsilon
_{ab}n_{(2)}^{a}\partial_{i}n_{(2)}^{b}+\frac{4q_{2}^{2}}{m}\left\vert
\psi_{2,t}\right\vert ^{2}\widetilde{\mathbf{A}}.
\end{equation}
By making use of $\phi-mapping$ topological current theory, the vorticity of
the current is given as%
\begin{equation}
\Gamma=\int_{\Sigma_{i}}\left(  \nabla\times\mathbf{J}\right)  \cdot
d\mathbf{S}=\int_{\Sigma_{i}}\left(  \nabla\times\mathbf{J}_{1}\right)  \cdot
d\mathbf{S+}\int_{\Sigma_{i}}\left(  \nabla\times\mathbf{J}_{2}\right)  \cdot
d\mathbf{S.}%
\end{equation}
Then the curl of the currents $\mathbf{J}_{1}$ and $\mathbf{J}_{2}$ are
calculated as
\begin{equation}
\left(  \nabla\times\mathbf{J}_{1}\right)  \mathbf{=}\frac{\hbar q_{1}\rho
_{t}^{\left(  1\right)  }}{m}%
{\displaystyle\sum\limits_{l=1}^{\ell}}
\beta_{l}^{\left(  1\right)  }\eta_{l}^{\left(  1\right)  }\delta_{\left(
1\right)  }^{2}\left(  x^{i}-z_{l}^{i}\right)  \frac{dx_{\left(  1\right)
}^{i}}{ds}+\frac{4q_{1}^{2}}{m}\left\vert \psi_{1,t}\right\vert ^{2}%
\mathbf{\nabla}\times\widetilde{\mathbf{A}}%
\end{equation}
and%
\begin{equation}
\left(  \nabla\times\mathbf{J}_{2}\right)  \mathbf{=}\frac{\hbar q_{2}\rho
_{t}^{\left(  2\right)  }}{m}%
{\displaystyle\sum\limits_{l=1}^{\ell}}
\beta_{l}^{\left(  2\right)  }\eta_{l}^{\left(  2\right)  }\delta_{\left(
2\right)  }^{2}\left(  x^{i}-z_{l}^{i}\right)  \frac{dx_{\left(  2\right)
}^{i}}{ds}+\frac{4q_{2}^{2}}{m}\left\vert \psi_{2,t}\right\vert ^{2}%
\mathbf{\nabla\times}\widetilde{\mathbf{A}}.
\end{equation}
Furthermore, the vorticity of the current is%
\begin{equation}
\Gamma=\frac{\hbar q_{1}\rho_{t}^{\left(  1\right)  }}{m}W_{1}+\frac{\hbar
q_{2}\rho_{t}^{\left(  2\right)  }}{m}W_{2}+\frac{4\left(  q_{1}^{2}\rho
_{t}^{\left(  1\right)  }+q_{1}^{2}\rho_{t}^{\left(  2\right)  }\right)  }%
{m}\int_{\Sigma_{i}}\left(  \mathbf{\nabla\times}\widetilde{\mathbf{A}%
}\right)  \cdot d\mathbf{S}.
\end{equation}
It is well known the property of a supercurrent is the magnetic flux passing
through any area bounded by such a current is quantized. The quantization of
the flux in the superconductor is%
\[
\int_{\Sigma_{i}}\left(  \mathbf{\nabla\times A}_{E}\right)  \cdot
d\mathbf{S=}\frac{h}{2e}\widetilde{W},
\]
where $e$ is the electronic charge. Similarly, we give the flux quantization
of this new $U\left(  1\right)  $ gauge potential $\widetilde{\mathbf{A}}$%
\begin{equation}
\int_{\Sigma_{i}}\left(  \mathbf{\nabla\times}\widetilde{\mathbf{A}}\right)
\cdot d\mathbf{S}=\frac{h}{q_{1}+q_{2}}\widetilde{W}.
\end{equation}
Finally, the vorticity of the current is%
\begin{equation}
\Gamma=\frac{\hbar q_{1}\rho_{t}^{\left(  1\right)  }}{m}W_{1}+\frac{\hbar
q_{2}\rho_{t}^{\left(  2\right)  }}{m}W_{2}+\frac{4\left(  q_{1}^{2}\rho
_{t}^{\left(  1\right)  }+q_{1}^{2}\rho_{t}^{\left(  2\right)  }\right)  }%
{m}\frac{\hbar}{q_{1}+q_{2}}\widetilde{W}.
\end{equation}

\section{conclusion}

In this paper, the relation between the coherent quantum system and the
charged two-condensate system is investigated. The new expression of the
current of the coherent state is given based on reduced density trajectory and
$\phi-mapping$ topological current theory. A topological reason leading to the
decoherence is found. By defining a new $U\left(  1\right)  $ gauge potential
$\widetilde{\mathbf{A}}$ and new charges $q_{1}$ and $q_{2},$ we find that the
coherent system can be described by the Ginzberg-Landau-like model with two
charged Cooper pairs. The corresponding relation between coherent system and
two-gap superconductor is shown as follows: the partial wave functions of the
coherence correspond to the charged two-condensate wave functions; the charges
$q_{1}$ and $q_{2}$ correspond to the electronic charges; the new $U\left(
1\right)  $ gauge potential $\widetilde{\mathbf{A}}$ corresponds to the
electromagnetic potential $\mathbf{A}e$. Finally, the hidden $O\left(
3\right)  $ symmetry of the coherent state is found using Faddeev's $O\left(
3\right)  $ nonlinear $\sigma$-model and the topological properties of the
knot solution are studied based on $\phi-$mapping topological current theory.

\begin{acknowledgments}
This work is supported by the Fundamental Research Funds for the Central Universities.
\end{acknowledgments}

\end{document}